\documentclass[twocolumn]{autart}    
\usepackage{graphicx}      
\usepackage{color}  
\usepackage{amsmath}
\usepackage{amssymb} 
\usepackage{mathtools}	
\usepackage{cite}
\usepackage{times}
\usepackage{enumitem}
\usepackage{subfigure}
\usepackage{hyperref}
\hypersetup{
	colorlinks=true,
	linkcolor=blue,
	citecolor=blue,
}

\newcommand{\chiup}{\raisebox{2pt}{$\chi$}}

\newcommand{\dist}{{\rm dist}}
\newcommand{\dt}{\frac{{\rm d}}{{\rm d}t}}
\newcommand{\defeq}{:=} 
\newcommand{\matr}[1]{\begin{bmatrix} #1 \end{bmatrix}}
\newcommand{\transpose}[1]{#1^\top}
\newcommand{\norm}[1]{\left\lVert#1\right\rVert}
\newcommand{\normm}[1]{\big\lVert#1\big\rVert}		
\newcommand{\tkp}{t_{k'}}
\newcommand{\tkpp}{t_{k''}}
\newcommand{\setp}{\mathcal{P}}
\newcommand{\setc}{\mathcal{C}}
\newcommand{\setm}{\mathcal{M}}
\newcommand{\mbr}[1][{}]{\mathbb{R}^{#1}}	
\newcommand{\cmmnt}[1]{}

\begin{document}

\begin{frontmatter}

\title{Path Following Control in 3D Using a Vector Field\thanksref{footnoteinfo}} 

\thanks[footnoteinfo]{Corresponding author: Weijia Yao. The work was supported in part by the European Research Council (ERC-CoG-771687) and the Netherlands Organization for Scientific Research (NWO-vidi-14134). Weijia Yao is funded by China Scholarship Council. }

\author[rug]{Weijia Yao}\ead{w.yao@rug.nl},             
\author[rug]{Ming Cao}\ead{m.cao@rug.nl}  
\address[rug]{ENTEG, University of Groningen, the Netherlands}


\begin{abstract}                          
	Using a designed vector field to control a mobile robot to follow a given desired path has found a range of practical applications, and it is in great need to further build a rigorous theory to guide its implementation. In this paper, we study the properties of a general 3D vector field for robotic path following. We stipulate and investigate assumptions that turn out to be crucial for this method, although they are rarely explicitly stated in the existing related works. We derive conditions under which the local path-following error vanishes exponentially in a sufficiently small neighborhood of the desired path, which is key to show the local input-to-state stability (local ISS) property of the path-following error dynamics. The local ISS property then justifies the control algorithm design for a fixed-wing aircraft model. Our approach is effective for any sufficiently smooth desired path in 3D, bounded or unbounded; the results are particularly relevant since unbounded desired paths have not been sufficiently discussed in the literature. Simulations are conducted to verify the theoretical results.
\end{abstract}

\end{frontmatter}

\section{Introduction} \label{sec1}
For path following, a control algorithm is designed to steer the output of a dynamical system to converge to and evolve along a pre-specified desired path. The desired path, usually not parametrized by time, is a geometric object which can be described by the zero-level set of an implicit smooth function \cite{Goncalves2010}. There are already many existing methods for path following\cite{belleter2019observer,consolini2010path,do2004robust}. Notably, it is shown in \cite{Sujit2014} that \textit{vector-field-based} path following algorithms achieve the smallest cross-track error while they require the least control efforts among several tested algorithms. In this context, a (guiding) vector field is carefully designed such that its integral curves are proven to converge to and traverse the desired path.

Many of the vector-field-based path following algorithms are only applicable to simple desired paths, such as circles and straight lines \cite{Nelson2007}. In addition, convergence to the desired path is often guaranteed locally in a small vicinity of the desired path. This is partly due to the fact that usually there are singular points\footnote{A point where a vector field becomes zero is called a \textit{singular point} of the vector field \cite[p. 219]{lee2015introduction}. The set of singular points of a vector field is called the \textit{singular set} of the vector field.} in the vector field. Recently, the work \cite{Y.A.2017} analyzes in details the properties of a 2D vector field for any desired path that is sufficiently smooth. The analysis of the singular points is also presented, and almost global convergence to the path is proven.

The previous study \cite{Y.A.2017}, as well as many other existing works \cite{Nelson2007,de2017guidance,Sujit2014}, only consider planar desired paths, while the 3D counterpart is less studied. In \cite{Goncalves2010}, given that the desired path is described by the intersection of several (hyper-)surfaces, a general vector field is proposed for robot navigation in the $n$-dimensional Euclidean space. However, strictly speaking, the analysis of this approach is only valid for (bounded) closed curves, such as circles, while the analysis cannot be directly applied to unbounded desired paths such as a straight line. Moreover, the assumption regarding the repulsiveness of the set of singular points is conservative. For example, this assumption is valid for a circle, but not for a Cassini oval or some other desired paths. In some literature, for ease of analysis, it is assumed that the workspace is free of singular points, but usually this is only guaranteed locally near the desired path. 

In this paper, we justify and employ a 3D guiding vector field for path following with rigorous analysis. Firstly, we present some general technical assumptions for vector-field-based path following, while relaxing the conservative assumption about the repulsiveness of singular points \cite{Goncalves2010}. Note that many existing related studies do not explicitly state these assumptions, while we point out that these assumptions are crucial in those reported results. Secondly, the convergence results and the maximal extensibility of solutions are analyzed rigorously. In addition, the conditions under which the local path-following error vanishes exponentially in a neighborhood of the desired path are provided, which is typically not available in the related literature. Thirdly, we show the local input-to-state stability of path-following error dynamics, which justifies the control algorithm design for a nonholonomic model: a fixed-wing aircraft. In comparison to many methods which only consider standard paths such as circles and straight lines, our method is applicable to any 3D desired path that can be described by the intersection of the zero-level sets of two implicit functions. And we specifically analyze rigorously the case of unbounded desired paths. Note that the analysis for the 3D vector field in this paper can be easily extended to any higher dimensional vector field.  Preliminary results are presented in our conference paper \cite{yao2018cdc}, in which we only prove the asymptotic convergence of the integral curves of the guiding vector field. In this paper, we show analysis of the local exponential convergence, the local ISS property and the normalization of the vector field, and present more comprehensive and elaborated technical details about the assumptions and the proofs for the case of unbounded desired paths.

Compared with \cite{Y.A.2017} and some other studies, we have provided additional analysis and features in this paper as follows:
	1) \textit{Our analysis of the 3D vector field can be directly applied to higher-dimensional vector fields after minor modifications.} 
	2) \textit{We have presented rigorous analysis for unbounded desired paths.} 
	3) \textit{We show the exponential vanishing of path-following error and the local ISS property.} 
	4) \textit{We elaborate on the assumptions that are crucial.}

The rest of this paper is organized as follows. Section \ref{sec2} presents the problem formulation. Then the analysis of the proposed vector field and its normalized and perturbed counterparts are elaborated in Section \ref{sec3} and \ref{sec4} respectively. The control algorithm for a fixed-wing aircraft model is provided in Section \ref{sec6}. Finally, Section \ref{sec7} concludes the paper.

\section{Problem Formulation} \label{sec2}
We first introduce some frequently-used concepts. Given a positive integer $n$, the distance between a point $p_0 \in \mathbb{R}^n$ and a set $\mathcal{S} \subset \mathbb{R}^n$ is denoted by $ \dist(p_0, \mathcal{S}) \defeq \inf\{ || p - p_0|| : p \in \mathcal{S} \}$. Similarly, the distance between two non-empty sets $\mathcal{A}$ and $\mathcal{B}$ is $\dist(\mathcal{A}, \mathcal{B}) =\dist(\mathcal{B}, \mathcal{A}) \defeq \inf\{ ||a-b||: a \in \mathcal{A}, b \in \mathcal{B} \}$. We say that a trajectory $\xi: [0, +\infty) \to \mathbb{R}^n$ asymptotically converges to a non-empty set $\mathcal{A} \subset \mathbb{R}^n$, if for any $\epsilon>0$, there exists $T>0$ such that $\dist(\xi(t),\mathcal{A}) < \epsilon$ for all $t > T$. For the case that the trajectory can only be maximally extended to $t_*<\infty$ \cite{khalil2002nonlinear}, we say that it converges to the nonempty set $\mathcal{A}$ as $t$ approaches $t_*$, if for any $\epsilon>0$, there exists $\delta>0$ such that $ \dist(\xi(t),\mathcal{A})<\epsilon$ for $|t-t_*| < \delta$.

Suppose the desired path $\setp$ is characterized by two functions $\phi_i:\mathbb{R}^3 \to \mathbb{R}, i=1,2,$ which are twice continuously differentiable:
\begin{equation} \label{path}
\mathcal{P} \defeq \{ \xi \in \mathbb{R}^3: \phi_1(\xi)=0, \phi_2(\xi)=0 \}.
\end{equation}
It is natural to assume that $\setp$ is nonempty, connected and one-dimensional. We will further require the regularity of the desired path as stated later in Assumption \ref{assump1}. One of the advantages of definition \eqref{path} is that the vector field can be derived directly from the function $\phi_i(\cdot)$ independent of the specific parametrization of the path. Another advantage is that the distance between a point $\xi \in \mbr[3]$ and the path $\dist(\xi, \setp) = \inf\{\norm{\xi - p}: p \in \setp \}$ can be approximated by the value of $\norm{(\phi_1(\xi),\phi_2(\xi))}$ under some assumptions presented later. Here, $\transpose{(\cdot)}$ denotes the transpose operation. 

We propose a 3D vector field $\chiup \in C^1: \mbr[3] \to \mbr[3]$ as follows:
\begin{equation} \label{3Dvf}
\chiup(\xi) = n_1(\xi) \times n_2(\xi) -  k_1 e_1(\xi) n_1(\xi) - k_2 e_2(\xi) n_2(\xi),
\end{equation}
where $n_i(\xi) = \nabla \phi_i(\xi)$ is the gradient of $\phi_i$, $k_i > 0$ are constant gains and the \textit{error function} $e_i= \phi_i(\xi)$ can be simply treated as the signed ``distance'' to the surfaces $\{\xi \in \mbr[3]: \phi_i(\xi)=0\}$ for $i=1,2$. For notational simplicity, we define $k_{{\rm min}}=\min\{k_1, k_2\}$ and $k_{{\rm max}}=\max\{k_1, k_2\}$ throughout the paper.
%
%
To write \eqref{3Dvf} in a compact form, let $ \tau(\xi) = n_1(\xi) \times n_2(\xi) $, $ N(\xi) = \big( n_1(\xi) , n_2(\xi) \big) $, $K= {\rm diag}(k_1,k_2)$ and $e(\xi)=\transpose{\big(e_1(\xi) , e_2(\xi)\big)}$. Then the vector field \eqref{3Dvf} is rewritten to
\begin{equation} \label{3Dvf2}
\chiup(\xi) =  \tau(\xi) -  N(\xi) K e(\xi).
\end{equation}
To study the properties of the vector field, we investigate the trajectory of the following nonlinear autonomous ordinary differential equation (ODE):
\begin{equation} \label{odegvf}
\dt \xi(t) = \chiup(\xi(t)), \quad t \ge 0.
\end{equation}
We aim to let the integral curves of the vector field converge to and move along the desired path. Namely, $\dist(\xi(t), \setp) \to 0$ as $t \to \infty$. Note that once the trajectory is on the desired path, the vector field degenerates to a set of \emph{tangent vectors} of the desired path (precisely, $\chiup(\xi)=\tau(\xi)$), thus the trajectory stays on the desired path and moves along it.

To carry out the analysis and to exclude some pathological cases, some assumptions are necessary. First we define the \textit{invariant set} $\mathcal{M}$ (its invariance will be shown later):
\begin{equation} \label{setinv}
\mathcal{M} = \{\xi \in \mathbb{R}^3 :   N(\xi) K e(\xi) = 0 \},
\end{equation}
and the \textit{singular set} $\mathcal{C}$: 
\begin{equation} \label{setcritical}
\mathcal{C} = \{\xi \in \mathbb{R}^3 : \chiup(\xi)=0 \} = \{ \xi \in \mathcal{M} : \text{rank}( N(\xi)) \le 1 \}. 
\end{equation}
The equivalence of the two expressions in \eqref{setcritical} can be seen as follows: if $n_1(\xi)$ and $n_2(\xi)$ are linearly independent, then they are also linearly independent with $n_1(\xi) \times n_2(\xi)$. Since the coefficient of $n_1(\xi) \times n_2(\xi)$ is non-zero, it is obvious that $ \chiup(\xi) \ne 0$. Therefore, the linear dependence of $n_1(\xi)$ and $n_2(\xi)$, which is equivalent to $\text{rank}(N(\xi)) \le 1$, is a necessary condition for $\chiup=0$. Also note that in the second expression, we restrict the elements to be in $\setm$. The elements of the singular set are \textit{singular points} of the vector field. Now we present the main assumptions in this paper.
\begin{assum} \label{assump1}
	It holds that $ \dist(\mathcal{P}, \mathcal{C}) > 0$.
\end{assum}
\begin{assum} \label{assump2}
	For any given constant $\kappa > 0$, we have
	$
	\inf\{ || e(\xi)||: \dist(\xi, \mathcal{P}) \ge \kappa\} > 0.
	$
\end{assum}
%
%
\begin{assum} \label{assump4}
	For any given constant $\kappa > 0$, we have 
	$
	\inf\{ || N(\xi) K e(\xi)||:  \dist(\xi, \mathcal{M}) \ge \kappa \} > 0. 
	$
\end{assum}
Assumption \ref{assump1} is needed for the ``regularity'' of the vector field. If there are singular points on the desired path, then a robot will get ``stuck'' on the desired path, since the ``translational velocity'' is zero at a singular point $c$ (i.e., $\tau(c)=0$).
	Assumption \ref{assump2} is motivated by observing that the desired path $\setp$ can be equivalently defined as 
	$
	\mathcal{P} = \{\xi \in \mathbb{R}^3 : e(\xi)=0 \}.
	$
	This inspires one to use $\norm{e(\xi)}$, the Euclidean norm of the vector function $e$, rather than the more complicated quantity $\dist(\xi, \setp)$, to quantify the distance between a point $\xi \in \mbr[3]$ and the desired path.  Note that although it is usually assumed that $\norm{e(\xi)}$ approximates the distance to the desired path $\dist(p, \mathcal{P})$, this is not always the case if Assumption \ref{assump2} is not verified. For instance, as the error $\norm{e(\xi)}$ converges to $0$, the distance $\dist(\xi, \mathcal{P})$ may even diverge to infinity (see Example 3 in \cite{yao2018cdc}). Thus, Assumption \ref{assump2} is crucial in the sense that it enables one to use a Lyapunov function candidate related to $\norm{e(\xi)}$ and the decreasing property to prove the convergence to the desired path conveniently.  A precise statement is presented in Appendix \ref{app2}. Therefore, under Assumption \ref{assump2}, we call $\norm{e(\xi)}$ the \textbf{path-following error}, or simply the \textbf{error}, of a point $\xi \in \mbr[3]$ to the desired path $\setp$ throughout the paper.
	Similarly, Assumption \ref{assump4} enables one to use $|| N(\xi) K e(\xi)||$ to measure the distance to the invariant set $\mathcal{M}=\setp \cup \setc$. It is suggestive to regard $|| N(\xi) K e(\xi)||$ as an ``error'' since when it equals $0$, the point $\xi$ is in $\mathcal{M}$, in view of the definition of $\mathcal{M}$ in \eqref{setinv}. Under Assumption \ref{assump4}, it can be similarly concluded that the vanishing of the ``invariant set error'' $|| N(\xi(t)) K e(\xi(t))||$ implies the convergence of the trajectory to the set $\mathcal{M}$, and hence to the desired path $\setp$ or the singular set $\setc$ exclusively as $t \to \infty$. A formal statement is in Appendix \ref{app2}.

\section{Analysis of the Vector Field} \label{sec3}

\subsection{Bounded Desired Path} 
Since the desired path is sufficiently smooth, a bounded desired path is the trace of a simple closed curve. It is proved below that the integral curves of \eqref{odegvf} asymptotically converge to either the desired path or the singular set. 
\begin{prop}\label{prop_bounded1}
	Let $\xi(t)$ be the solution of \eqref{odegvf}. If the desired path $\mathcal{P}$ is bounded, then $\xi(t)$ will asymptotically converge to the desired path or the singular set exclusively as $t \rightarrow \infty$.
\end{prop}
\begin{pf}[Sketch of proof] The full proof is presented in \cite{yao2018cdc}. The main idea is using the LaSalle's invariance theorem \cite[Theorem 4.4]{khalil2002nonlinear}. 	We define a Lyapunov function candidate by 
	\begin{equation} \label{eqlyapunov}
	V(\xi(t))= 1/2 \; \transpose{e}(\xi(t)) K e(\xi(t)),
	\end{equation} 
	Taking the derivative of $V$ with respect to $t$, we obtain $\dot{V}(\xi(t)) = -||N(\xi(t)) K e(\xi(t)) ||^2 \le 0$ on $\mathbb{R}^3 \setminus \mathcal{P}$. To use LaSalle's invariance theorem, we need to construct a compact set that is positively invariant with respect to \eqref{odegvf}. Given $r > 0$, a (closed) ball is defined by $\overline{\mathcal{B}}_r=\{ \xi\in\mathbb{R}^3 : ||\xi|| \le r \} \subset \mathbb{R}^3$. Since $\mathcal{P}$ is bounded, $r$ can be chosen sufficiently large such that 
	$\mathcal{P} \subset \overline{\mathcal{B}}_r$,  and $\alpha = \min_{||\xi||=r} V(e(\xi)) > 0$. Take $\beta \in (0, \alpha) $ and let
	\begin{equation} \label{setomegabeta}
	\Omega_\beta = \{ \xi \in \overline{\mathcal{B}}_r : V(e(\xi)) \le \beta \}.
	\end{equation}
	Obviously, $\Omega_\beta$ is in the interior of $\overline{\mathcal{B}}_r$, and hence it is compact. In addition, since $\dot{V}(\xi(t)) \le 0$, the set $\Omega_\beta$ is also positively invariant. By the LaSalle's invariance theorem \cite[Theorem 4.4]{khalil2002nonlinear}, the solution will converge to the set $\setm$, and thus  converge either to the desired path or the singular set as $t \rightarrow \infty$ by Lemma \ref{lemma_converge2pc}. \hfill$\qed$
\end{pf}
Asymptotic convergence is not quite appealing compared to exponential convergence. For this reason, we show as follows the exponential convergence result. We will use the Lyapunov function candidate in \eqref{eqlyapunov}, and the compact set $\Omega_\beta$  in the proof of Proposition \ref{prop_bounded1}, which is proved to be positively invariant. Moreover, we define two more sets:
\begin{equation} \label{setneighbor}
\mathcal{E}_\alpha=\{\xi \in \mathbb{R}^3 : \norm{e(\xi)} \le \alpha\},
\end{equation}
which is the set of points at which the error is less than some positive number $\alpha$. This set can be treated as the (closed) neighborhood of the desired path $\setp$. Another set is defined by
\begin{equation} \label{setcprime}
\mathcal{C'}=\{\xi \in \mbr[3] \setminus \setm : \tau(\xi) = 0 \}.
\end{equation}
This set contains all the points $\xi \in \mathbb{R}^3 \setminus \setm$ where the gradient vectors $n_1(\xi)$ and $n_2(\xi)$ are linearly dependent (including the case where either of them is zero). Thus $\setc \cap \mathcal{C'} = \emptyset$. Now the theorem is stated below:

\begin{thm} \label{thm_localexp1}
	Let $\xi(t)$ be the solution to \eqref{odegvf} and suppose that the desired path $\mathcal{P}$ is bounded. If $\dist(\setp, \mathcal{C'})>0$, where $\mathcal{C}'$ is defined in \eqref{setcprime}, then there exists $\delta>0$ such that $\mathcal{E}_{\delta}$ defined in \eqref{setneighbor} is compact, and $\norm{\tau(\xi)}\ne0$ for every point $\xi \in \mathcal{E}_{\delta}$. Furthermore, the error $\norm{e(\xi(t))}$ (locally) exponentially converges to $0$ as $t \rightarrow \infty$, given that the initial condition $\xi(0) \in \mathcal{E}_{\delta'}$, where $0<\delta'\le \delta \sqrt{k_{\rm min}/k_{\rm max}}$.
\end{thm}
\begin{pf}
	Since $K$ is positive definite, from \eqref{eqlyapunov}, we have
	$
	\norm{e(\xi)}^2 \ge 2 V(e(\xi)) / k_{{\rm max}} .
	$
	Taking the derivative of \eqref{eqlyapunov} with respect to time, we have ($t$ is omitted for simplicity):
	$
	\dot{V}(e(\xi)) =  -\transpose{e}(\xi) Q(\xi) e(\xi) = -\norm{N(\xi) K e(\xi)}^2,
	$
	where
	\begin{equation} \label{eqq}
	Q(\xi) = K \transpose{N}(\xi) N(\xi) K 
	\end{equation}
	is positive semidefinite. Note that $\det(Q(\xi)) = k_1^2 k_2^2 \norm{\tau(\xi)}^2 $. Therefore, $\det(Q(\xi)) \ne 0$ if and only if $n_1$ and $n_2$ are linearly independent. 
	Under Assumption $\ref{assump1}$ (i.e., $\dist(\setp, \setc)>0$) and the condition that $\dist(\setp, \mathcal{C'})>0$, for any point $\xi \in \setc \cup \mathcal{C'}$, we have $\dist(\xi, \setp)>0$ and $\norm{\tau(\xi)}=0$, and hence $\norm{e(\xi)} \ge \gamma$ for some positive number $\gamma$ (Assumption \ref{assump2}). Therefore, there exists $0<\delta \le \gamma$, such that for any point $\xi \in \mathcal{E}_\delta$ as defined in \eqref{setneighbor}, we have $\norm{\tau(\xi)} \ne 0$. Note that $\delta$ can be chosen sufficiently small such that $\mathcal{E}_\delta$ is bounded, hence compact\footnote{This is justified as follows: one can choose a set $\Omega_\beta$ as defined in \eqref{setomegabeta}, which is compact. Then there exists $\gamma' > 0$ such that $\mathcal{E}_{\gamma'} \subset \Omega_\beta$ (this is true because by choosing $\gamma' \le \sqrt{2 \beta / k_{{\rm max}}}$, $\forall \xi \in \mathcal{E}_{\gamma'}, \norm{e(\xi)} \le \gamma' \implies V(\xi) \le k_{{\rm max}} \norm{e(\xi)}^2 / 2 \le k_{{\rm max}} \gamma'^2 / 2 \le \beta \implies \xi \in \Omega_\beta$). Therefore, $\mathcal{E}_{\gamma'}$ is compact. Finally, by selecting $0<\delta<\min\{\gamma, \gamma'\}$, it can be guaranteed that $\mathcal{E}_\delta$ is compact as desired (since $\mathcal{E}_\delta \subset \mathcal{E}_{\gamma'} \subset \Omega_\beta$).}. 
	Let $\iota = k_{{\rm min}} \delta^2 /2$, then $\Omega_\iota \subset \mathcal{E}_\delta$\footnotemark. \footnotetext{Since $\forall \xi \in \Omega_\iota, k_{{\rm min}} \norm{e(\xi)}^2 / 2 \le V(e(\xi)) \le \iota \implies \norm{e(\xi)} \le \delta \implies \xi \in \mathcal{E}_\delta$.} Therefore, in the compact and positively invariant set $\Omega_\iota$, we have $\norm{\tau(\xi)} \ne 0$, implying that $Q(\xi)$ does not loose rank, and further implying that $Q(\xi)$ is positive definite. Let $\Lambda \defeq \min_{\xi \in \Omega_\iota} \{\lambda_{\rm min}(Q(\xi)) \}$, where $\lambda_{\rm min}(\cdot)$ denotes the minimum eigenvalue. It can be observed that $\Lambda > 0$. Note that $\Lambda$ always exists because the eigenvalues of a matrix continuously depends on its entries, and the minimum is obtained over a compact set. Therefore,
	$
	\dot{V}(e(\xi)) \le - \Lambda \norm{e(\xi)}^2 \le -2\Lambda V(e(\xi)) / k_{{\rm max}} ,
	$
	which implies that
	$
	V(e(\xi)) \le V(e_0) \exp \left( {-2 \Lambda t / k_{{\rm max}}  } \right),
	$
	and furthermore,
	$
	||e(\xi)|| \le c ||e_0|| \exp \left( {-\Lambda t / k_{{\rm max}} } \right), 
	$
	where $e_0 = e(\xi(0))$ and $c=\sqrt{k_{{\rm max}}/k_{{\rm min}}}$. Therefore, $\norm{e(\xi(t))}$ exponentially approaches $0$ as $t$ approaches infinity.  Lastly, note that $\mathcal{E}_{\delta'} \subset \Omega_\iota$; thus $\xi(0) \in \mathcal{E}_{\delta'} \implies \xi(0) \in \Omega_\iota$.   \hfill$\qed$
\end{pf}
\begin{rem} \label{remark13}
	For a 2D vector field as proposed  in \cite{Y.A.2017}, the set $\mathcal{C}' = \emptyset$, and thus the assumption $\dist(\setp, \mathcal{C'})>0$ in Theorem \ref{thm_localexp1} is straightforwardly true. However, for the three-dimensional case, it is possible that $\mathcal{C}' \ne \emptyset$ and thus the problem becomes more complicated. This situation also appears for higher-dimensional vector fields. 
\end{rem}

\subsection{Unbounded Desired Path} 

The analysis presented above for bounded desired paths cannot be directly applied to an unbounded desired path. This is partly because for any (closed) ball $\overline{\mathcal{B}}_r$ containing part of the desired path, $\alpha = \min_{||\xi||=r} V(e(\xi)) = 0$. Therefore, $\beta \in (0, \alpha)$ is not valid in the definition of $\Omega_\beta$ in \eqref{setomegabeta}. The key issue is that LaSalle's invariance theorem is no longer effective regarding an unbounded desired path, since a compact set containing the desired path is not possible. In addition, the solution to \eqref{odegvf} may not be extended infinitely. Therefore, we need to analyze this case differently. 
\subsubsection{Extensibility of Solutions}
Assuming that $||\tau||=||n_1 \times n_2||$ is upper bounded on some set, it can still be proved that the solution can be extended to infinity. We consider the following unbounded set:
\begin{equation} \label{eqxibeta}
\Xi_\beta = \{ \xi \in \mathbb{R}^3 : V(e(\xi)) \le \beta \},
\end{equation}
where $\beta>0$. The definition is similar to that of $\Omega_\beta$ in \eqref{setomegabeta}, except that $\Xi_\beta$ is unbounded since $\setp \subset \Xi_\beta$ and $\setp$ is unbounded.
\begin{lem} \label{inftime}
	Suppose $||\tau||$ is upper bounded in $\mathcal{E}_\alpha$ defined in \eqref{setneighbor} for some $\alpha>0$. Let $\xi(t)$ be the trajectory w.r.t. \eqref{odegvf}. If the initial condition $\xi(0) \in \mathcal{E}_{\alpha'}$, where $0<\alpha'\le \alpha \sqrt{k_{\rm min}/k_{\rm max}}$, then the trajectory $\xi(t)$ can be extended to infinity.
\end{lem}
\begin{pf}
	Suppose the maximum extended time $t^*$ of the solution is finite; i.e., $t^*<\infty$. Let $\beta = k_{{\rm min}} \alpha^2/2$. First one observes that $\Xi_\beta \subset \mathcal{E}_\alpha$ (since $\forall x \in \Xi_\beta, k_{{\rm min}} \norm{x}^2 / 2 \le V(x) \le \beta \implies \norm{x} \le \sqrt{2 \beta / k_{{\rm min}}} \le \alpha \implies x \in \mathcal{E}_\alpha$). Using the same Lyapunov function as in \eqref{eqlyapunov}, its derivative with respect to $t$ is $\dot{V}(e(\xi(t)))=-||N(\xi(t)) K e(\xi(t)) ||^2 \le 0$. Therefore, $\Xi_\beta$ is positively invariant (note that in this case $\Xi_\beta$ is not bounded). This means that $\xi(0) \in \Xi_\beta \implies \xi(t) \in \Xi_\beta \subset \mathcal{E}_\alpha$ for $t \in [0, t^*)$, where $\xi(t)$ is the trajectory with respect to \eqref{odegvf}. In other words, $\norm{\tau(\xi(t))}$ is upper bounded by some positive number denoted by $\kappa_b$ for all $t \in [0, t^*)$.
	Since $V(\xi(t)) \ge 0$, it follows that	
	$
	\int_{0}^{t^*} ||N(\xi(t)) K e(\xi(t)) ||^2 dt 
	= -\int_{0}^{t^*} \dot{V}(\xi(t)) dt =  V(\xi(0)) - V(\xi(t^*)) < \infty. 
	$
	Therefore, for all $0 \le \tilde{t}<t^*$, 
	$
	||\xi(\tilde{t})-\xi(0)|| \le \int_{0}^{\tilde{t}}|| \dot{\xi}(t)|| dt  \le   \int_{0}^{\tilde{t}} ||\tau(\xi(t))|| dt +  \int_{0}^{\tilde{t}} ||N(\xi(t)) K e(\xi(t)) || dt \le  \kappa_b t^* + \sqrt{t^* \int_{0}^{t^*} ||N(\xi(t)) K e(\xi(t)) ||^2 dt } \defeq R < \infty. 
	$
	The last inequality is due to H\"{o}lder's inequality. Therefore the trajectory $\xi(t)$ remains in a compact set  $\{p \in \mathbb{R}^3 : ||p - \xi(0)|| \le R \}$, and hence the trajectory can be extended to infinity. Lastly, note that $\mathcal{E}_{\alpha'} \subset \Xi_\beta$; thus $\xi(0) \in \mathcal{E}_{\alpha'} \implies \xi(0) \in \Xi_\beta$.	 \hfill$\qed$
\end{pf}

\begin{cor} \label{coroll3}
	Suppose the assumptions of Lemma \ref{inftime} are satisfied. Then along the trajectory $\xi(t)$ w.r.t. \eqref{odegvf}, we have
	\begin{equation} \label{eq14}
	\int_{0}^{\infty} ||N(\xi(t)) K e(\xi(t)) ||^2 dt = -\int_{0}^{\infty} \dot{V}(\xi(t)) dt < +\infty.
	\end{equation}
\end{cor}

\subsubsection{Convergence Results}
We can draw a similar conclusion for the case of an unbounded desired path. To this end, we present the absolute continuity of the Lebesgue integral first.
\begin{lem}[Absolute continuity of Lebesgue integrals \cite{jones2001lebesgue}] \label{lemlebes}
	If $f$ is Lebesgue integrable on $\mathbb{R}^n$, then for any $\epsilon>0$, there exists $\delta>0$ such that for all measurable sets $\mathcal{D} \subset \mathbb{R}^n$ with measure $m(\mathcal{D})<\delta$, it follows that
	$
	\int_{\mathcal{D}}^{} |f| dm < \epsilon.
	$
\end{lem}
Now we are ready to prove the following result.
\begin{cor} \label{coroll4}
	For any $\epsilon>0$, there exists $0<\delta \le \epsilon$ such that for all intervals with length $|\Delta|<\delta$, 
	$
	\int_{\Delta}^{{}} || N(\xi(t)) K e(\xi(t))|| dt < 2 \epsilon.
	$
\end{cor}
\begin{pf}
	For any function $f: \mathbb{R} \rightarrow \mathbb{R}^n$, we define a new function $f(t)_{>1}=f(t), \forall ||f(t)||>1$ and $f(t)_{>1}=0$, otherwise.
	Another function $f(t)_{\le 1}$ is similarly defined. It follows that
	$
	\int_{0}^{\infty} || N(\xi(t)) K e(\xi(t))||_{>1} dt 
	\le \int_{0}^{\infty} || N(\xi(t)) K e(\xi(t))||^2_{>1} dt 
	\le \int_{0}^{\infty} || N(\xi(t)) K e(\xi(t))||^2 dt \\ < \infty,
	$
	where the last inequality is due to Corollary \ref{coroll3}. Therefore, $|| N(\xi(t)) K e(\xi(t))||_{>1}$ is Lebesgue integrable. Thus, for any $\epsilon>0$, there exists $\gamma>0$ as the length of the interval such that Lemma \ref{lemlebes} holds. In addition, taking $\delta=\min\{\gamma, \epsilon\}$, then $|\Delta|$ can be chosen sufficiently small such that $|\Delta|< \delta \le \epsilon$, and $\int_{\Delta}^{{}} || N(\xi(t)) K e(\xi(t))||_{>1} dt < \epsilon$. Finally,
	$
	\int_{\Delta}^{{}} || N(\xi(t)) K e(\xi(t))|| dt 
	= \int_{\Delta}^{{}} || N(\xi(t)) K e(\xi(t))||_{>1} dt + \int_{\Delta}^{{}} || N(\xi(t)) K e(\xi(t))||_{\le 1} dt \le \epsilon + \epsilon = 2 \epsilon.
	$ \hfill$\qed$
\end{pf}
The following proposition for an unbounded desired path is the counterpart of Proposition \ref{prop_bounded1}. 
\begin{prop} \label{prop_unbound}
	Let $\xi(t)$ be the solution of \eqref{odegvf}. If $\mathcal{P}$ is unbounded and the assumptions of Lemma \ref{inftime} are satisfied, then the trajectory $ \xi(t)$ will asymptotically converge to the desired path or the singular set exclusively as $t \to \infty$.
\end{prop}
\begin{pf}
	Define the Lyapunov function candidate $V(\xi(t))$ as in Proposition \ref{prop_bounded1} and denote $\eta(\xi(t))=||N(\xi(t)) K e(\xi(t))||$. Suppose $\xi(t)$ does not converge to $\mathcal{M}$, then there exists a sequence $\{ t_k \}$, and $t_k \rightarrow \infty$ as $k \rightarrow \infty$, such that (due to Assumption \ref{assump4})
	$
	\dist(  \xi(t_k), \mathcal{M}) > \delta > 0 \Rightarrow  \eta(\xi(t_k))  > \epsilon > 0.
	$
	Therefore, $\dot{V}(  \xi(t_k)) = -\eta^2(\xi(t_k)) < -\epsilon^2$. According to Assumption \ref{assump4}, there exists $\epsilon' > 0$ such that when $\dist(\xi,\mathcal{M}) > \delta/2 $, one has $|| \eta(\xi)|| > \epsilon'$. Since $\dist(\xi(t_k), \mathcal{M}) > \delta$, given a ball $\mathcal{B}(\xi(t_k), \delta/4)$, then for any $y \in \mathcal{B}(\xi(t_k), \delta/4)$, it follows that 
	$
	\dist(y, \mathcal{M}) > \delta/2 \Rightarrow \dot{V}(y)  < -\epsilon'^2.
	$
	Taking $\epsilon=\delta / (2(2+\kappa_b))$ in Corollary \ref{coroll4}, then there exists an interval $\Delta$ with length $|\Delta| < \epsilon$ such that
	$
	\int_{\Delta}^{{}} ||\dot{\xi}(t)|| dt = \int_{\Delta}^{{}}  ||\tau(\xi(t)) -  N(\xi(t)) K e(\xi(t)) || dt 
	\le \int_{\Delta}^{{}} ||\tau(\xi(t))|| dt + \int_{\Delta}^{{}} || N(\xi(t)) K e(\xi(t))|| dt 
	\le (\kappa_b + 2) \epsilon < \delta / 2.
	$
	Then it follows that
	$
	\xi [t_k - \Delta/2, t_k + \Delta/2] \subset \mathcal{B}(  \xi (t_k), \delta/4).
	$
	Therefore, 
	$
	\int_{t_k-\Delta/2}^{t_k+\Delta/2} \dot{V}(\xi (t)) dt < -\epsilon'^2 \Delta.
	$
	This leads to
	$
	\int_{0}^{\infty} \dot{V}(\xi(t)) dt \le \sum_{k=1}^{\infty} \int_{t_k - \Delta/2}^{t_k + \Delta/2} \dot{V}(\xi(t)) dt 
	\le - \sum_{k=1}^{\infty}  \epsilon'^2 \Delta \le -\infty,
	$
	which contradicts Corollary \ref{coroll3}. Therefore,  $  \xi(t)$ converges to $\mathcal{M}$ as $t \rightarrow \infty$. Then due to Assumption \ref{assump1}, the solution converges either to the desired path or the singular set.	 \hfill$\qed$
\end{pf}



For unbounded desired paths, we also have the following exponential convergence result. Before presenting the result, we say that a function $f:\Omega \subset \mbr[m] \to \mathbb{R}^n$ is bounded away from zero in $\Omega$ if there exists a real number $c>0$, such that $\norm{f(x)}>c$ for all $x \in \Omega$.
\begin{thm} \label{thm_localexp2}
	Let $\xi(t)$ be the solution to \eqref{odegvf} and the desired path $\mathcal{P}$ be unbounded. Define $\mathcal{E}_\alpha$ as in \eqref{setneighbor} for some $\alpha>0$. Suppose both $\norm{n_1(\xi)}$ and $\norm{n_2(\xi)}$ are upper bounded in $\mathcal{E}_\alpha$, and $\norm{\tau(\xi)}$ is bounded away from zero on $\setp$, then there exists $0<\gamma\le\alpha$ such that $\inf_{\xi \in \mathcal{E}_\gamma}  \norm{\tau(\xi)} > 0$. Furthermore, the error $\norm{e(\xi)}$ (locally) exponentially converges to $0$ as $t \rightarrow \infty$, given that the initial condition $\xi(0) \in \mathcal{E}_{\gamma'}$, where $0<\gamma'\le \gamma \sqrt{k_{\rm min}/k_{\rm max}}$.
\end{thm}
\begin{pf}
	It is obvious that the assumptions of Lemma \ref{inftime} are satisfied. Thus the solution $\xi(t)$ w.r.t \eqref{odegvf} can be prolonged to infinity. Since the desired path is unbounded, we cannot find a compact set $\Omega_\beta$ as in Theorem \ref{thm_localexp1}. Instead, we consider $\Xi_\beta$ defined in \eqref{eqxibeta}. Since $\norm{\tau}$ is bounded away from zero on $\setp$, and due to the continuity of $\tau(\xi)$ with respect to its argument, there exists $0 < \gamma \le \alpha$ such that $\inf_{\xi \in \mathcal{E}_\gamma}  \norm{\tau(\xi)} > 0$. That is, $\norm{\tau}$ is bounded away from zero in the subset $\mathcal{E}_\gamma \subset \mathcal{E}_\alpha$. It can be shown that there exists a positively invariant set $\Xi_\beta \subset \mathcal{E}_\gamma$ by choosing $\beta=k_{\rm min}\gamma^2/2$ (see the proof in Lemma \ref{inftime}),  where $\Xi_\beta$ is defined in \eqref{eqxibeta}. Next we consider the case where the solution $\xi(t)$ starts from this invariant set $\Xi_\beta$. Since $\norm{\tau}$ is bounded away from zero in the subset $\mathcal{E}_\gamma \supset \Xi_\beta$ as shown previously, there are no singular points in $\Xi_\beta$, and thus we do not need to consider the case where the solution converges to the singular set, and thus the remaining proof is similar to that of Theorem \ref{thm_localexp1}. It follows that 
	$
	\inf_{\xi \in \Xi_\beta} \lambda_1(Q(\xi)) \lambda_2(Q(\xi)) = \inf_{\xi \in \Xi_\beta} \det(Q(\xi)) = k_1^2 k_2^2 \inf_{\xi \in \Xi_\beta}  \norm{\tau(\xi)}^2 > 0
	$,
	where $\lambda_1(Q(\xi))$ and $\lambda_2(Q(\xi))$ are two eigenvalues of $Q(\xi)$. Note that the sum of the two eigenvalues $\lambda_1(Q(\xi)) + \lambda_2(Q(\xi)) = {\rm tr}(Q(\xi)) = k_1^2 \norm{n_1}^2 + k_2^2 \norm{n_2}^2$. Since $\norm{n_1}$ and $\norm{n_2}$ are upper bounded in $\Xi_\beta \subset \mathcal{E}_\alpha$, the two eigenvalues are finite. Therefore, we have $\Lambda' \defeq \inf_{\xi \in \Xi_\beta} \{ \lambda_{\rm min}(Q(\xi)) \} > 0$. 
	This leads to
	$
	\dot{V}(e(\xi)) \le -\Lambda' \norm{e(\xi)}^2 \le 2 \Lambda' V(e(\xi)) / k_{{\rm max}} .
	$
	Therefore,
	$
	V(e(\xi)) \le V(e_0) \exp \left( {- 2 \Lambda' t / k_{{\rm max}}  } \right) \notag	 
	\implies ||e(\xi)|| \le c ||e_0|| \exp \left( {- \Lambda' t / k_{{\rm max}} } \right), 
	$
	where $e_0 = e(\xi(0))$ and $c=\sqrt{k_{{\rm max}}/k_{{\rm min}}}$. Therefore, the error $\norm{e(\xi)}$ will exponentially approach $0$ as $t$ approaches infinity.  Lastly, note that $\mathcal{E}_{\gamma'} \subset \Xi_\beta$; thus $\xi(0) \in \mathcal{E}_{\gamma'} \implies \xi(0) \in \Xi_\beta$.  \hfill$\qed$
\end{pf}

\begin{rem}
	For an unbounded desired path, the result presented above is valid under the condition that $||\tau||=\norm{n_1 \times n_2}$ is upper bounded. This seems restrictive. However, a smooth bouding operator $f_b: \mathbb{R}^n \rightarrow \mathbb{R}^n$, $||f_b(\tau)||$ may be introduced \cite{yao2018cdc}. Nevertheless, for practical reasons, it is desirable to normalize the original vector field, while compromising the maximal extensibility of the solutions. This will be discussed in the next section.
\end{rem}

\section{Normalization and Perturbation of the Vector Field} \label{sec4}
In this section, based on the results presented above, we study the properties of a normalized 3D vector field. We show that the essential feature of the vector field is the direction rather than the amplitude at each point in $\mbr[3]$. Then the robustness of the vector field against perturbation is also analyzed.

%
For notational simplicity, we define the \textit{normalization operator} $\hat{\cdot}: \mbr[n] \to \mbr[n]$ which normalizes a given nonzero vector $a$ such that $\hat{a} \defeq a / \norm{a}$. Therefore, the desired direction of velocity at location $\xi \in \mathbb{R}^3$ is represented by 
$
\hat{\chiup}(\xi),
$
where $\chiup(\xi)$ is the vector field in \eqref{3Dvf}. This vector field is well defined in the open set $\mathbb{R}^3 \setminus \setc$, where $||\chiup|| \ne 0$.  The integral curves of the normalized vector field correspond to the solution to the following autonomous ODE:
\begin{equation} \label{odegvf2}
\dt \xi(t) = \hat{\chiup}(\xi(t)), 
\end{equation}
where $\xi: \mathbb{R}_{\ge0} \to \mathbb{R}^3 \setminus \setc$. The existence and uniqueness of solutions of the ODE can be guaranteed since the right-hand side of \eqref{odegvf2} is continuously differentiable in $\mathbb{R}^3 \setminus \setc$. 
Note that the vector field in \eqref{odegvf2} differs from that in \eqref{odegvf} by a positive scalar function that only depends on the states $\xi$. Therefore, these two vector fields have the same direction of each vector at the same point. This fact implies that there is a bijection between non-equilibrium solutions of the two differential equations \eqref{odegvf} and \eqref{odegvf2}. Recall that a \emph{phase portrait} or \emph{phase diagram} is a geometric picture of all the orbits of an autonomous differential equation \cite[p. 9]{chicone2006ordinary}.

\begin{lem} \label{lemma_reparam}
	The ODE \eqref{odegvf2} with a normalized vector field and the ODE \eqref{odegvf} with the original vector field have the same phase portrait in $\mathbb{R}^3 \setminus \setc$.
\end{lem}
\begin{pf}
	The right-hand side of \eqref{odegvf2} can be written as $ \hat{\chiup}(\xi) = \chiup(\xi) / ||\chiup(\xi)|| $, where the original vector field $\chiup(\xi)$ is scaled down by a positive and continuously differentiable function $1 / ||\chiup(\xi)||$ in $\mathbb{R}^3 \setminus \setc$. Therefore, the ODE \eqref{odegvf2} with a normalized vector field is obtained from the ODE \eqref{odegvf} by a re-parametrization of time \cite[Proposition 1.14]{chicone2006ordinary}. Therefore, they have the same phase portrait in $\mathbb{R}^3 \setminus \setc$ \cite{chicone2006ordinary}.  \hfill$\qed$
\end{pf}
Since the differential equation \eqref{odegvf2} is defined in $\mathbb{R}^3 \setminus \setc$, the maximal interval to which a solution can be extended is finite when the solution is approaching $\setc$.
%
\begin{lem} \label{lemma_finitetime}
	Let $\xi(t)$ be a solution to \eqref{odegvf2}. If the solution is maximally extended to $t^* < \infty$, then it will converge to the singular set; that is, $ \lim_{t\rightarrow t^*}  \dist(\xi(t), \mathcal{C})=0$.
\end{lem}
\begin{pf}
	Since   $\normm{\dot{\xi}(t)}$ is bounded, $\xi^*:=\lim_{t \rightarrow t^*} \xi(t) = \xi(0) + \int_{0}^{t^*}\dot{\xi}(t) dt$ exists. To show that $ ||\chiup(\xi^*)|| = 0$, suppose $ ||\chiup(\xi^*)|| > 0$. Since $\chiup$ continuously depends on $\xi$, the same holds in the vicinity of $\xi^*$, and hence the right-hand side of \eqref{odegvf2} is well defined and bounded in the vicinity of $\xi^*$. This enables one to define the solution at $t=t^*$ and, by the existence theorem \cite{khalil2002nonlinear}, extend to $[t^*, t^*+\epsilon)$ for some $\epsilon>0$. We arrive at the contradiction with the definition of $t^*$, which proves that $ ||\chiup(\xi^*)|| = 0$. Thus the solution will converge to the singular set.  \hfill$\qed$
\end{pf}
Due to Lemma \ref{lemma_finitetime}, the solution to \eqref{odegvf2} will possibly converge to the singular set in finite time. However, it can still be similarly proved that the trajectory will either converge to the desired path or the singular set by Lemma \ref{lemma_reparam}. Furthermore, the exponential convergence results still hold under the conditions of Theorem \ref{thm_localexp1} for bounded desired paths or Theorem \ref{thm_localexp2} for unbounded desired paths. The corresponding results are not presented due to page limits.

Now we consider a system with a perturbed vector field
\begin{equation} \label{eqperturb}
\dot{\xi}(t) = \chiup(\xi(t)) + d(t),
\end{equation}
where $\chiup$ is the vector field in \eqref{3Dvf} and $d: \mbr[]_{\ge 0} \to \mbr[n]$ is a piecewice continuous and bounded function of time $t$ for all $t\ge0$. Therefore, the dynamics for the path-following error w.r.t. to \eqref{eqperturb} is
\begin{equation} \label{eqerrdyn}
\dot{e}(t) = \transpose{N(\xi(t))} (\chiup(\xi(t)) + d(t)).
\end{equation}
It will be proved subsequently that the path-following error dynamics \eqref{eqerrdyn} is locally input-to-state stable \cite{khalil2002nonlinear,sontag1995characterizations}. We will use the definition of an open ball: given $a>0$, the open ball $\mathcal{B}_a \subset \mbr[n]$ is defined as $\mathcal{B}_a \defeq \{\xi \in \mbr[n] : \norm{\xi} < a\}$. 
\begin{thm} \label{thm_liss}
	Suppose that the desired path $\mathcal{P}$ is bounded and $\dist(\setp, \mathcal{C'})>0$, where $\mathcal{C}'$ is defined in \eqref{setcprime}. Then the path-following error \eqref{eqerrdyn} is locally input-to-state stable. 
\end{thm}
\begin{pf}
	From Theorem \ref{thm_localexp1}, there exists $\delta>0$ such that $\mathcal{E}_{\delta}$ defined in \eqref{setneighbor} is compact, and $\norm{\tau(\xi)}\ne0$ for every point $\xi \in \mathcal{E}_{\delta}$, and thus the eigenvalue 
		$
		\Lambda' \defeq \min_{\xi \in \mathcal{E}_{\delta}} \{\lambda_{\rm min}(Q(\xi)) \}>0,
		$
		where the matrix $Q$ is defined in \eqref{eqq}.	We use the same Lyapunov function in \eqref{eqlyapunov} and take the time derivative:
		\begin{align}
		\dot{V} &= - \norm{N K e}^2 + \transpose{d} N K e \\
		&\le -\frac{1}{2} \norm{N K e}^2 + \frac{1}{2} \norm{d}^2 \label{eqine1} \\ 
		&\stackrel{\eqref{eqq}}{=} -\frac{1}{2} \transpose{e} Q e + \frac{1}{2} \norm{d}^2 \\
		&\le -\frac{1}{2} \Lambda' \norm{e}^2 + \frac{1}{2} \norm{d}^2 \label{eqine2} \\
		& \le -\frac{\epsilon}{2} \Lambda' \norm{e}^2, 	\quad \forall \norm{e} \ge \rho(\norm{d})>0, \label{eqine3}
		\end{align}
		for all $(t,e,d) \in [0, \infty) \times \mathcal{B}_{\delta} \times \mathcal{B}_{r}$, where $r=\delta \sqrt{(1-\epsilon)\Lambda'}$ with $0<\epsilon<1$, and $\rho(\norm{d})=\norm{d}/\sqrt{(1-\epsilon)\Lambda'}$ is a class $\mathcal{K}$ function.
		Note that \eqref{eqine1} is due to Young's inequality (i.e., $\transpose{d} N K e \le \norm{d}^2/2 + \norm{NKe}^2/2$). Also note that \eqref{eqine2} is verified since we have restricted $e \in \mathcal{B}_\delta$. The disturbance is also restricted to $d \in \mathcal{B}_r$ such that $\rho(\norm{d})<\delta$ is satisfied and \eqref{eqine3} is valid. Therefore, the path-following error in \eqref{eqerrdyn} is locally input-to-state stable by the \emph{local version} of Theorem 4.19 in \cite{khalil2002nonlinear}.  \hfill$\qed$
\end{pf}
\begin{rem}
		This theorem indicates that the error satisfies 
		$
		\norm{e(\xi(t))} \le \beta(\norm{e(\xi(0))}, t) + \gamma \left( \sup_{s \in [0,t]} \norm{d(s)} \right)
		$
		for a class $\mathcal{K}\mathcal{L}$ function $\beta$ and a class $\mathcal{K}$ function $\gamma$. If the disturbance $d(\cdot)$ is vanishing, then the error $\norm{e(\xi(t)} \to 0$ as $t \to \infty$; if the disturbance $d(\cdot)$ is non-vanishing but bounded, then the error $\norm{e(\xi(t))}$ will be uniformly ultimately bounded by a class $\mathcal{K}$ function of $\sup_{s \in [0,\infty)} \norm{d(s)}$. 
\end{rem}
\begin{rem}
	This theorem can be easily adapted for unbounded desired paths if the assumptions of Theorem \ref{thm_localexp2} are satisfied. The significance of this theorem is that it justifies the design of control algorithms: one can focus on designing a control algorithm such that the direction of the robot's velocity converges to that of the vector field. 
\end{rem}
\section{Control Algorithm for a Fixed-wing Aircraft} \label{sec6}
We use the following fixed-wing aircraft kinematic model discussed in \cite{rezende2018}:
\begin{subequations} \label{model}
	\begin{align}
	\dot{x} &= s \cos \theta \label{modela} \\
	\dot{y} &= s \sin \theta \label{modelb} \\
	\dot{z} &= \tau_z^{-1} (-z + z_u) \label{modeld} \\
	\dot{\theta} &= \tau_\theta^{-1} (-\theta + \theta_u) \label{modelc} \\
	\dot{s} &= \tau_s^{-1} (-s + s_u), \label{modele}
	\end{align}
\end{subequations}
where $(x,y,z)$ is the position of the center of mass of the aircraft, $s>0$ is the airspeed, $\theta$ is the yaw angle, $\tau_z>0$, $\tau_\theta>0$ and $\tau_s>0$ are the time constants, and $z_u$, $\theta_u$ and $s_u$ are the control inputs. The control of $z$ coordinate in \eqref{modeld} and the airspeed $s$ in \eqref{modele} are independent from the other variables. 
Therefore, we can first consider the planar orientation control. Denote the orientation of the aircraft on the $X$-$Y$ plane and that of the normalized vector field $\hat{\chiup}$ on the $X$-$Y$ plane by $h^p(\theta)$ and $\chiup^p$ respectively; that is,
$
h^p(\theta) \defeq \transpose{(\cos\theta , \sin\theta)}$
and
$
\chiup^p \defeq \transpose{(\hat{\chiup}_1 , \hat{\chiup}_2)},
$
where $\hat{\chiup}_1$ and $\hat{\chiup}_2$ are the first two entries of $\hat{\chiup}$. Note that the superscript $p$ implies that the vector is the projection on the $X$-$Y$ plane. To utilize the vector field designed and analyzed before, it is desirable that $h^p$ is steered to align with $\chiup^p$. In other words, we want to achieve $\hat{h^p} \to \hat{\chiup^p}$, where $\hat{\cdot}$ is the normalization operator defined before. For convenience, we call $\hat{h^p}$ and $\hat{\chiup^p}$ the \textit{planar orientations} of the aircraft and of the vector field respectively. It can be observed that $\hat{h^p} = h^p$ and 
$
\hat{\chiup^p} = \chiup^p / \norm{\chiup^p} = \transpose{(\hat{\chiup}_1 , \hat{\chiup}_2)}/\sqrt{\hat{\chiup}_1^2 + \hat{\chiup}_2^2} = \transpose{(\chiup_1 , \chiup_2)}/\sqrt{ \chiup_1^2 + \chiup_2^2}.
$
The following theorem gives the angle control input $\theta_u$ which can steer the planar orientation of the aircraft to that of the vector field asymptotically.
\begin{thm}
	Let the angle directed from $\hat{\chiup^p}$ to $\hat{h^p}$ be denoted by $\beta \in (-\pi, \pi]$. When the control input in \eqref{modelc} takes the form
		\begin{align} 
		\theta_u &= \tau_\theta (\dot{\theta_d} - k_\theta \transpose{\hat{h^p}} E \hat{\chiup^p}) + \theta,	\label{eqthetau} \\
		\dot{\theta_d} &= \frac{-1}{\norm{\chiup^p}} \transpose{\hat{\chiup^p}} E J(\chiup^p) \dot{\xi}, \label{eqthetad}
		\end{align}
		where $E=\matr{0 & -1 \\ 1 & 0}$ is the \textit{rotation matrix} of angle $\pi/2$, $k_\theta$ is a positive gain, $\dot{\xi}=(\dot{x},\dot{y},\dot{z})$ is the aircraft's actual velocity and $J(\chiup^p)$ is the Jacobian matrix of $\chiup^p$ with respect to $\xi$, then the angle $\beta(t) \to 0$ as $t \to \infty$ whenever $\beta(0) \in (-\pi, \pi)$. 
\end{thm}
\begin{pf}
	Substituting \eqref{eqthetau} into \eqref{modelc}, one has
		\begin{equation} \label{eqthetadot}
		\dot{\theta}=\dot{\theta_d} - k_\theta \transpose{\hat{h^p}} E \hat{\chiup^p}.
		\end{equation}
		First, one can calculate that 
		$
		\dt \hat{h^p} = \transpose{(-\sin\theta , \cos\theta)} \dot{\theta} = \dot{\theta} E \hat{h^p},
		$
		and
		$
		\dt \hat{\chiup^p} = ( -\transpose{\hat{\chiup^p}} E J(\chiup^p) \dot{\xi} / \norm{\chiup^p} ) E \hat{\chiup^p} = \dot{\theta_d} E \hat{\chiup^p}.
		$
		Note that
		$
		\cos\beta = \transpose{\hat{h^p}} \hat{\chiup^p}.
		$
		Taking the time derivative of both sides of the previous equation, we have:
		$
		-\sin\beta \cdot \dot{\beta} = \transpose{(\dot{\theta} E \hat{h^p})} \hat{\chiup^p} + \transpose{\hat{h^p}} \dot{\theta_d} E \hat{\chiup^p} 
		= (\dot{\theta_d} - \dot{\theta}) \transpose{\hat{h^p}} E \hat{\chiup^p} 
		\stackrel{\eqref{eqthetadot}}{=} k_\theta (\transpose{\hat{h^p}} E \hat{\chiup^p} )^2 
		= k_\theta \sin^2 \beta
		$,
		where the last equality is due to $\transpose{\hat{h^p}} E \hat{\chiup^p}=\sin\beta$. Therefore, the dynamics of the angle $\beta$ is simply
		$
		\dot{\beta} = -k_\theta \sin\beta.
		$
		Since $\beta \in (-\pi, \pi]$, there are two equilibria $\beta=0$ and $\beta=\pi$ in the previous differential equation. Using linearization \cite[Theorem 4.7]{khalil2002nonlinear}, it is easily shown that the equilibrium $\beta=\pi$ is unstable while the other equilibrium $\beta=0$ is asymptotically stable. One also observes that $\dot{\beta}<0$ when $\beta \in (0, \pi)$ and $\dot{\beta}>0$ when $\beta \in (-\pi,0)$. Therefore, whenever $\beta(0) \in (-\pi, \pi)$, the trajectory of the angle $\beta(t)$ will asymptotically converge to $0$, inferring that $\hat{h^p} \to \hat{\chiup^p}$ asymptotically as $t \to \infty$.	 \hfill$\qed$
\end{pf}
This theorem implies that the planar orientation of the robot $\hat{h^p}$ will asymptotically converge to that of the vector field $\hat{\chi^p}$ (i.e., $\hat{h^p} \to \hat{\chi^p}$) almost globally with respect to the initial angle difference $\beta(0)$. The altitude and airspeed control are more straightforward. Since the planar orientation of the aircraft $\hat{h^p}=\transpose{(\dot{x} , \dot{y})}/s$ will approach that of the vector field $\hat{\chiup^p}$ using the control input $\theta_u$ developed in the previous part, it is desirable that $\dot{z}$ equals the third component of the vector field $\chiup_3$. However, in view of \eqref{modela} and \eqref{modelb}, since $\norm{(\dot{x}, \dot{y})}=s$, $\dot{z}$ should be scaled accordingly to $\dot{z} = s \, \chiup_3 / \sqrt{\chiup_1^2 + \chiup_2^2}$. Therefore, from \eqref{modelc} it can be computed that the altitude control input is
\begin{equation} \label{eqzu}
z_u = z + \tau_z\, s\, \chiup_3 / \sqrt{\chiup_1^2 + \chiup_2^2}.
\end{equation}
The idea of scaling is the same as that in \cite{rezende2018}. Next, to let the aircraft fly at the constant nominal speed (cruise speed) $s^*$, the airspeed control input in \eqref{modele} is simply
\begin{equation} \label{eqsu}
s_u = s^*.
\end{equation}
Therefore, the control input $\theta_u, z_u$ and $s_u$ result in the orientation difference between the aircraft and the 3D vector field asymptotically vanishes. If this orientation error is regarded as a vanishing disturbance, then according to the local ISS property in Theorem \ref{thm_liss}, the path-following error will also vanish, and thus path following behavior is successfully realized.

\section{Simulations}
\begin{figure}
	\centering
	\includegraphics[width=0.9\linewidth]{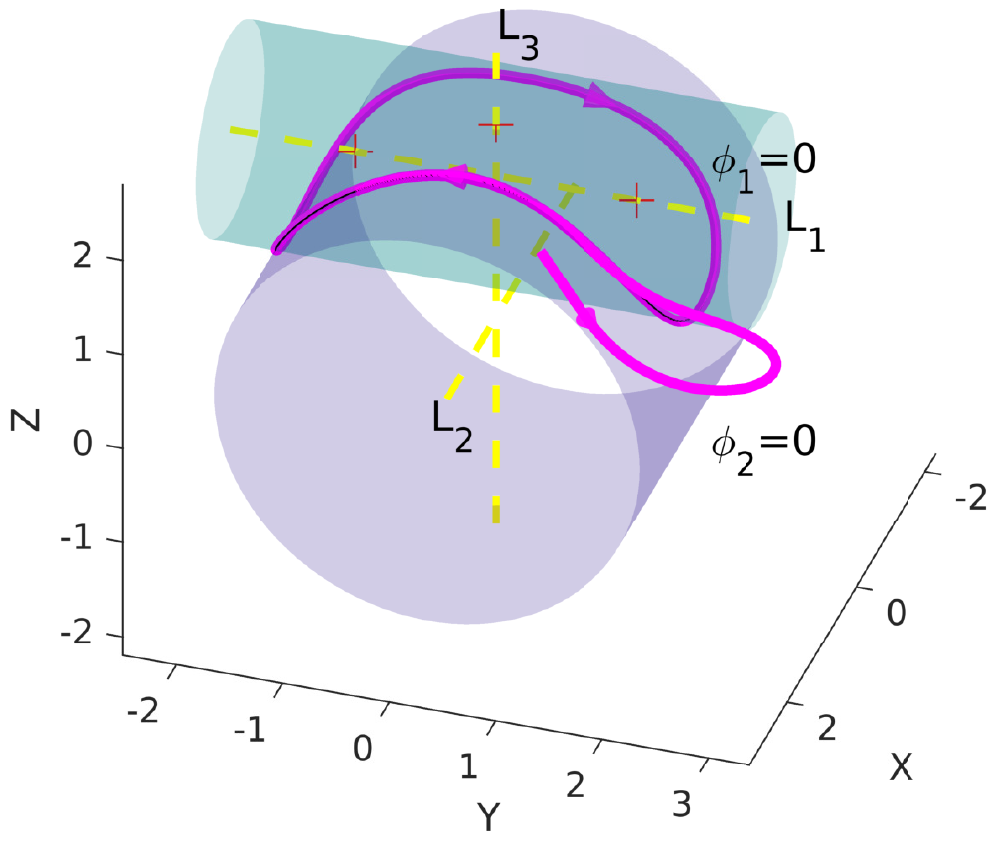}
	\caption{The fixed-wing aircraft successfully follows a 3D bounded desired path. The actual trajectory and the desired path overlaps. The arrows indicate the orientation of the aircraft.}
	\label{fig:boundedtraj}
\end{figure}

\begin{figure}
	\centering
	\includegraphics[width=0.8\linewidth]{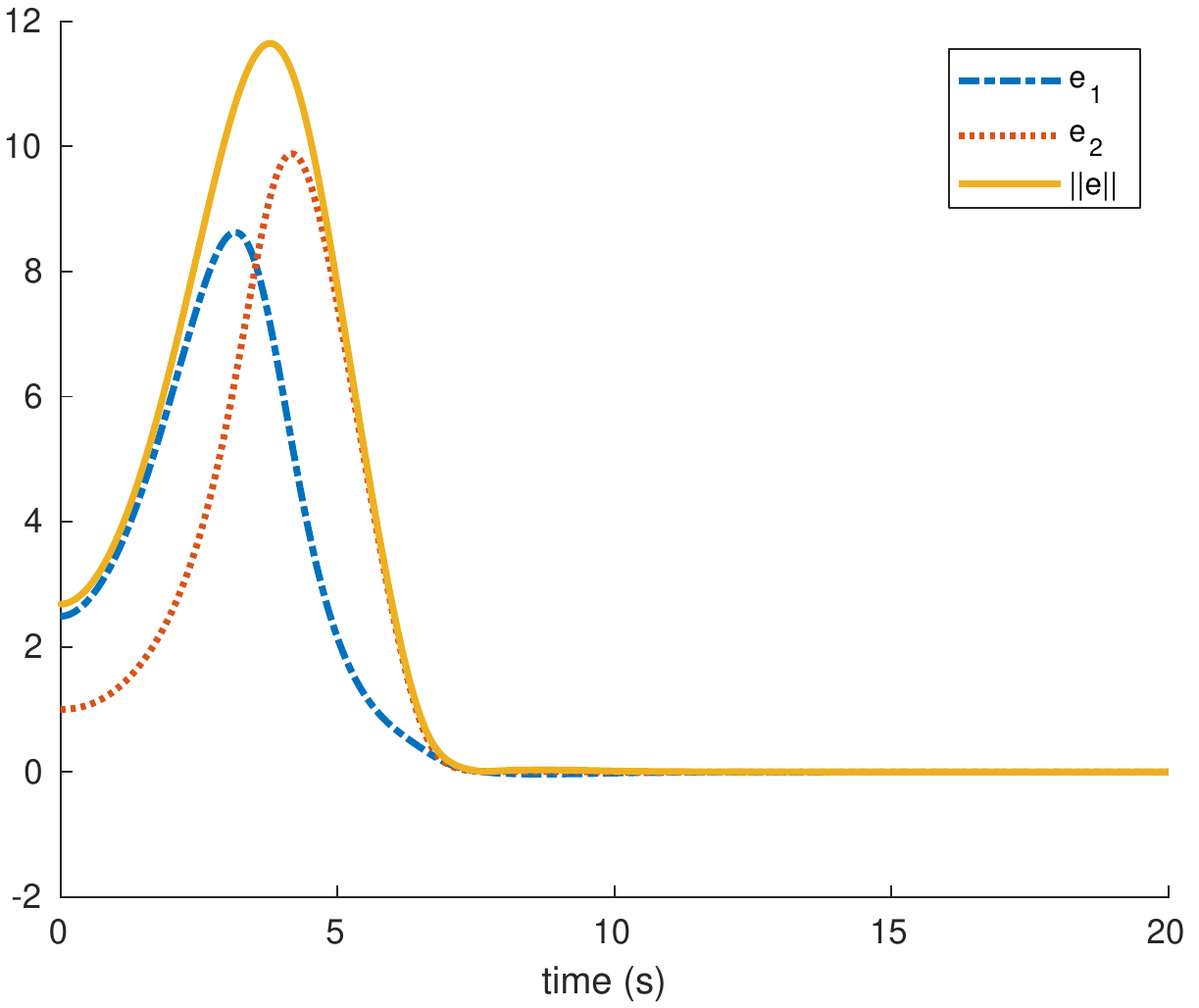}
	\caption{The path-following errors for the first simulation.}
	\label{fig:boundederror}
\end{figure}

The first simulation considers a bounded desired path in 3D. It is described by the intersection of two cylindrical surfaces, $\phi_1=0$ and $\phi_2=0$. Specifically,
\[
\phi_1(\xi) = (x-a)^2 + (z-b)^2 - r^2,	\quad \phi_2(\xi) = y^2 + z^2 - R^2,
\]
where $a, b, R, r \in \mbr[]$ are parameters. We choose $R=2$, $r=1$, $a=0$, $b=1.5$. The desired path is shown in Fig. \ref{fig:boundedtraj}. Then the vector field $\chi(\xi)$ is readily obtained according to \eqref{3Dvf} with $k_1=k_2=2$. It can be calculated that there are only three isolated singular points in this vector field (cross marks in Fig. \ref{fig:boundedtraj}). In addition, the set $\mathcal{C}'$ in \eqref{setcprime} is $\mathcal{C}' = (L_1 \cup L_2 \cup L_3) \setminus \mathcal{C}$, where $L_1$, $L_2$ and $L_3$ are three straight lines shown as dashed lines in Fig. \ref{fig:boundedtraj}. To be more precise, $L_1=\{(a, y, b) \in \mbr[3] : y \in \mbr[]\}$, $L_2=\{(x, 0, 0) \in \mbr[3] : x \in \mbr[]\}$ and $L_3=\{(a, 0, z) \in \mbr[3] : z \in \mbr[]\}$. Therefore, the control inputs in \eqref{eqthetau}, \eqref{eqzu} and \eqref{eqsu} are used to guide the aircraft to follow this path. The other parameters are:  $\tau_z=\tau_\theta=\tau_s=1$, $k_\theta=1$ and $s^*=1$. The initial value of the kinematics model \eqref{model} is $(x(0), y(0), z(0), \theta(0), s(0)) = (1.8, 1, 2, \pi/4, 0)$. The aircraft trajectory is the solid line shown in Fig. \ref{fig:boundedtraj}, and the error $\norm{e}$ is plotted in Fig. \ref{fig:boundederror}. As can be seen from the figure, the error $\norm{e}$ is not monotonically decreasing. The initial increase of the error is due to the fact that the robot cannot move in any arbitrary direction; it first needs to steer its orientation towards that of the vector field, resulting in movement further away from the desired path in the beginning (see the beginning segment of the trajectory in Fig. \ref{fig:boundedtraj}). However, the aircraft successfully follows the desired bounded path as the error eventually converges to zero. 

\begin{figure}
	\centering
	\includegraphics[width=0.9\linewidth]{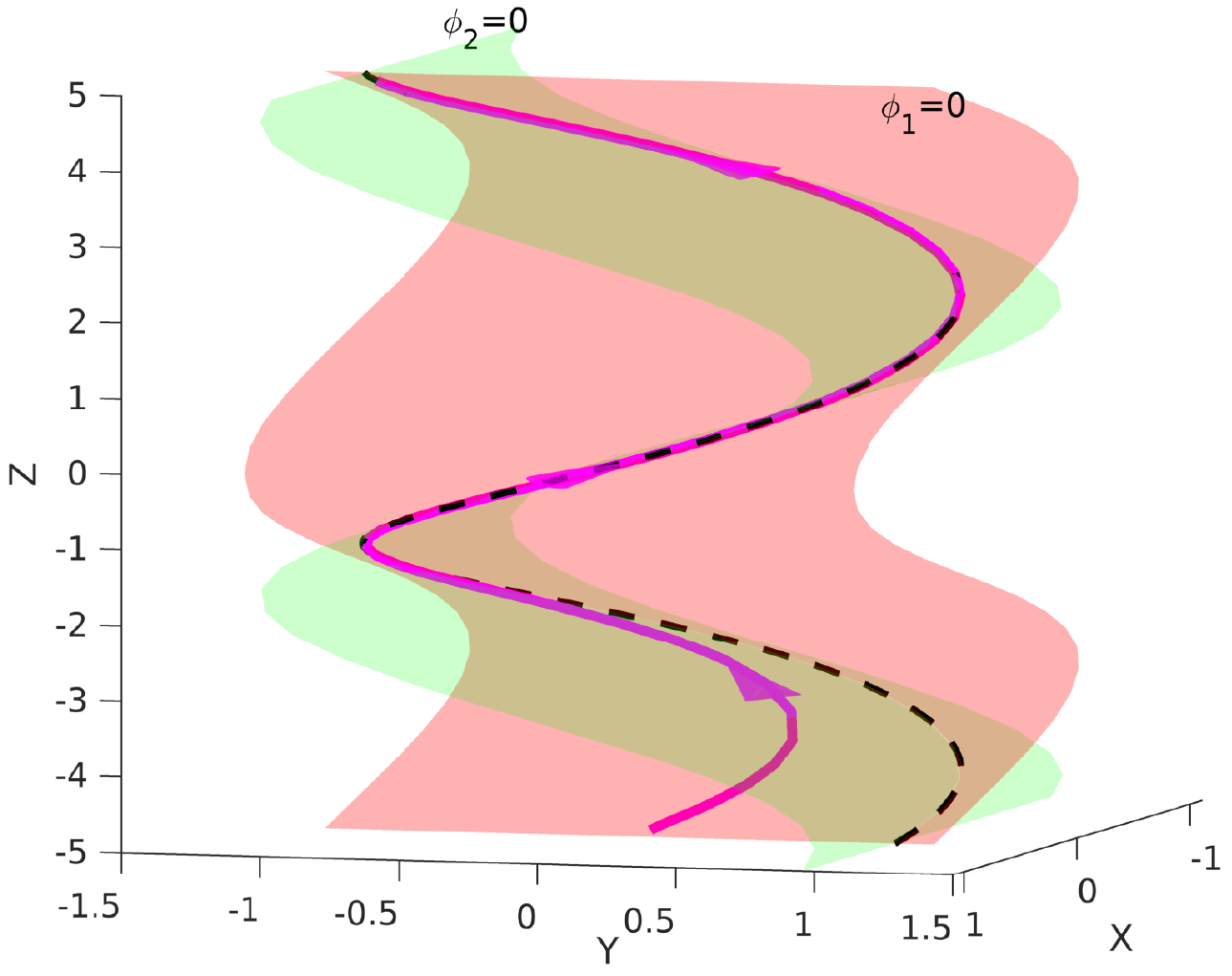}
	\caption{The trajectory of the fixed-wing aircraft (the solid line) gradually overlaps the 3D unbounded desired path (the dashed line). The arrows indicate the orientation of the aircraft.}
	\label{fig:unboundedtraj}
\end{figure}

For the 3D unbounded path, we choose a helix described by
\[
\phi_1(\xi) = x - \cos z,	\quad \phi_2(\xi) = y - \sin z.
\]
It can be easily calculated that $n_1 = \transpose{(1, 0, \sin z)}$, $n_2 = \transpose{(0, 1, -\cos z)}$ and $\tau=n_1 \times n_2 = \transpose{(-\sin z, \cos z, 1)}$. It is interesting to note that there are no singular points in this case as $\tau \ne 0$ in $\mbr[3]$. In addition, since $\norm{n_1} \le \sqrt{2}$, $\norm{n_2} \le \sqrt{2}$ and $\norm{\tau}= \sqrt{2}$, the assumptions in Theorem \ref{thm_localexp2} are satisfied (globally). Therefore, the control inputs in \eqref{eqthetau}, \eqref{eqzu} and \eqref{eqsu} can be used to guide the aircraft to follow this path. The other parameters are: $\tau_z=\tau_\theta=\tau_s=1$, $k_1=k_2=k_\theta=1$ and $s^*=1$. The initial value is $(x(0), y(0), z(0), \theta(0), s(0)) = (0.1, 0, -5, \pi, 0)$. The aircraft trajectory is the solid line shown in Fig. \ref{fig:unboundedtraj}, and the error $\norm{e}$ is plotted in Fig. \ref{fig:boundederror}. As can be seen from the figures, the aircraft successfully follows the desired unbounded path.
\begin{figure}
	\centering
	\includegraphics[width=0.8\linewidth]{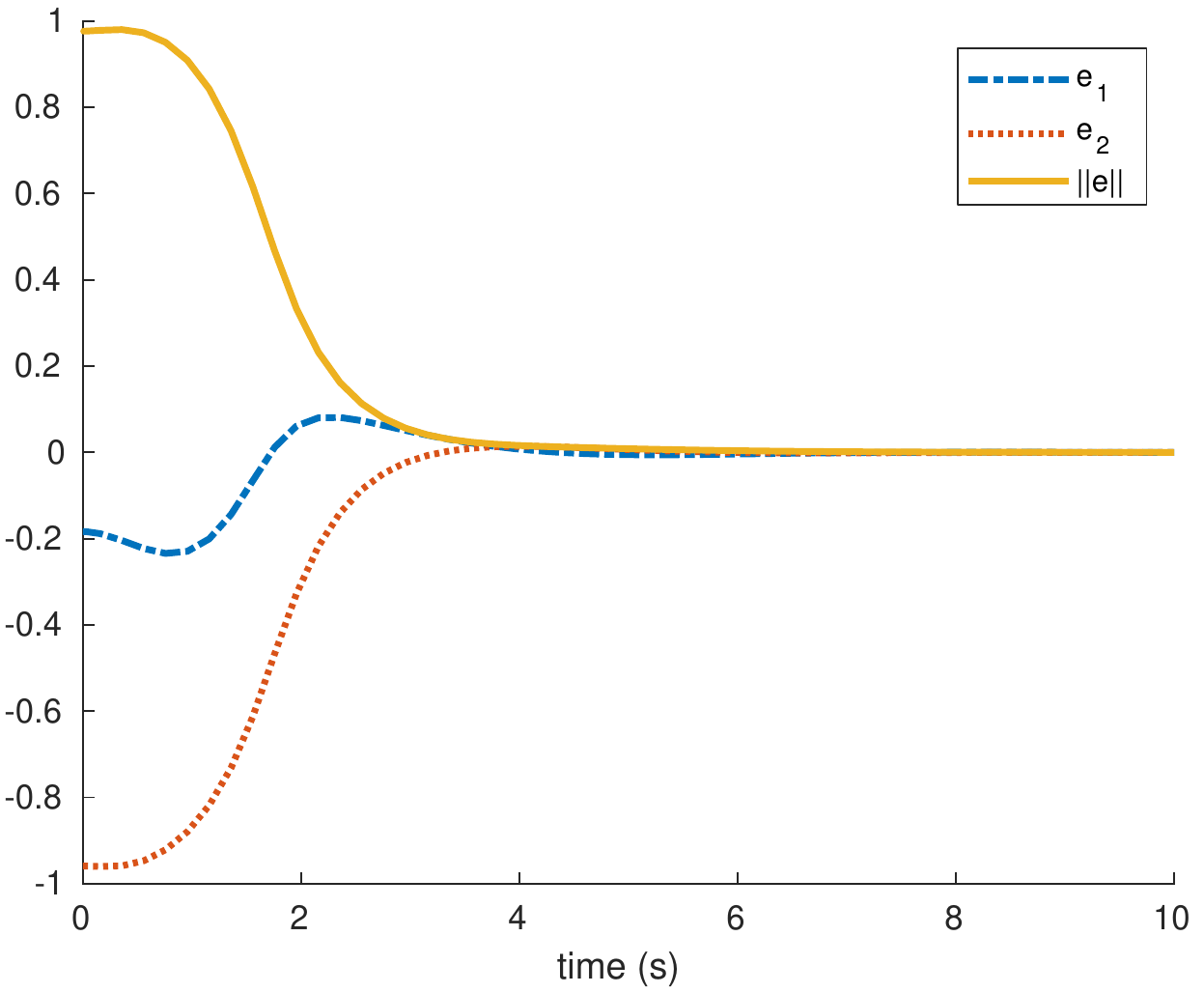}
	\caption{The path-following errors for the second simulation.}
	\label{fig:unboundederror}
\end{figure}
\section{Conclusion} \label{sec7}
We have provided rigorous theoretical results for path following control using a 3D vector field. Crucial assumptions are presented and elaborated. Based on this, we have shown the asymptotic and exponential convergence of the path-following error for both bounded and unbounded desired paths. Furthermore, the local ISS property of the path-following error dynamics is proved, which justifies the control algorithm designed for a nonholonomic model: a fixed-wing aircraft. Our vector field method is flexible in the sense that it is valid for any general desired path that is sufficiently smooth, and its extension to higher-dimension is straightforward.
We are interested in designing a new control algorithm to mitigate the effect of wind on the path following performance of a fixed-wing aircraft. 

\section*{Acknowledgement}
We would like to thank the reviewers for their valuable comments which improve the paper. We are also thankful to Bohuan Lin and Yuri A. Kapitanyuk for valuable discussions.

\appendix
\section{Appendix} \label{app2}

Assumption \ref{assump2} is utilized to avoid the pathological situation where the distance $\dist(\xi, \mathcal{P})$ diverges to infinity as the error $\norm{e(\xi)}$ converges to $0$ (see Example $3$ in \cite{yao2018cdc}). More specifically, Assumption \ref{assump2} leads to the following lemma. 
\begin{lem} \label{lemma_errorconverge}
	Let $ \big( p(t_k) \big)_{k=1}^{\infty} $ be a sequence of points in $\mathbb{R}^3$, where $t_k$ is a strictly increasing sequence of $k$ and $t_k \rightarrow \infty$ as $k \rightarrow \infty$. Under Assumption \ref{assump2}, if the error converges to zero, then $\big( p(t_k) \big)_{k=1}^{\infty}$ converges to the desired path $\mathcal{P}$. Equivalently, 
	\[
	\lim_{k \rightarrow \infty} ||e(p(t_k))||=0 \implies \lim_{k \rightarrow \infty}  \dist(p(t_k), \mathcal{P})=0.
	\]
\end{lem}
\begin{pf}
	We prove this lemma by contradiction. If $\big( p(t_k) \big)_{k=1}^{\infty}$ does not converge to the desired path $\mathcal{P}$, then
	\begin{equation} \label{eq8}
	(\exists \epsilon>0) (\forall L>0)(\exists k' \ge L) \; \dist(p(t_{k'}), \mathcal{P}) \ge \epsilon.
	\end{equation} 
	According to Assumption \ref{assump2}, replacing $\kappa$ by $\epsilon$, we suppose
	\begin{equation} \label{eq9}
	\inf\{ || e(p)||:  \dist(p, \mathcal{P}) \ge \epsilon \} = \beta > 0.
	\end{equation}
	Since  $\lim_{k \rightarrow \infty} ||e(p(t_k))||=0$, it follows that
	\begin{equation} \label{eq10}
	(\exists L' >0) (\forall k \ge L') \; \norm{e(p(t_k))} < \beta.
	\end{equation}
	Let $L$ be chosen as $L'$ in \eqref{eq8}, then there exists $k' \ge L'$ such that $\dist(p(t_{k'}), \mathcal{P}) \ge \epsilon$. Then due to \eqref{eq9}, $\norm{e(p(t_k))} \ge \beta$, which contradicts \eqref{eq10}. Therefore, $\big( p(t_k) \big)_{k=1}^{\infty}$ indeed converges to the desired path $\mathcal{P}$.	\hfill$\qed$
\end{pf}

Due to Assumption \ref{assump1} and Assumption \ref{assump4}, the convergence to the set $\mathcal{M}$ implies that the trajectory converges exclusively to the desired path $\mathcal{P}$ or the singular set $\mathcal{C}$. This is stated in the following lemma.
\begin{lem} \label{lemma_converge2pc}
	Let $p: \mathbb{R}_+ \to \mathbb{R}^3$ be a continuous function. If $\setp, \setc \ne \emptyset$, $\dist(\setp, \setc) = \beta > 0$ (i.e., Assumption \ref{assump1}), and $p(t)$ converges to $\mathcal{M}$ as $t \to \infty$,  then $p(t)$ converges to either $\mathcal{P}$ or $\mathcal{C}$ exclusively as $t \to \infty$. 
\end{lem}

Before the proof of Lemma \ref{lemma_converge2pc}, we first present a lemma which is similar to the triangle inequality (the proof is neglected due to its simplicity).
\begin{lem}[Triangle inequality for $\dist$] \label{lemma_triangle}
	Let $\mathcal{A}, \mathcal{B} \subset \mathbb{R}^n$ be two non-empty sets. Suppose $\dist(\mathcal{A}, \mathcal{B}) = \beta \ge 0$, then for any point $p \in \mathbb{R}^n$, $\dist(p, \mathcal{A}) + \dist(p, \mathcal{B}) \ge \beta$.
\end{lem}
%
%
\begin{pf}[Proof of Lemma \ref{lemma_converge2pc}]
	Suppose $\dist(\mathcal{P}, \mathcal{C})=\beta>0$. First, it is obvious that $p(t)$ does not converge to both $\mathcal{P}$ and $\mathcal{C}$ at the same time. 
	%
	\begin{figure}[htb] 
		\centering
		\includegraphics[width=0.5\linewidth]{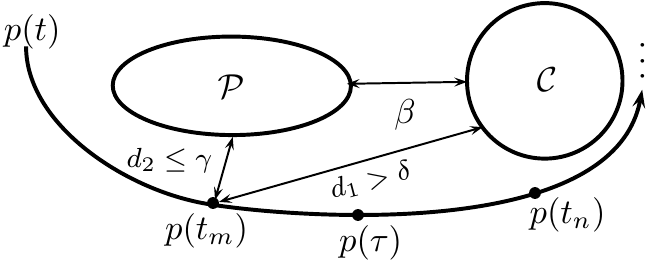}
		\caption{Proof of Lemma \ref{lemma_converge2pc}. In this figure, $d_1=\dist(p(t_m), \setc) > \delta$, $d_2=\dist(p(t_m), \setp) \le \gamma < \delta$, $\beta$ is the distance between $\setp$ and $\setc$.}
		\label{fig:lemma_convergepc}
	\end{figure}
	Next we show that $p(t)$ does converge to either $\mathcal{P}$ or $\mathcal{C}$. We prove this by contradiction. Suppose $p(t)$ converges neither to $\mathcal{P}$ nor to $\mathcal{C}$. Then there exist $\delta>0$, and two subsequences $\big( p(\tkp) \big)_{k'=1}^{\infty}$ and $\big( p(\tkpp) \big)_{k''=1}^{\infty}$, where $\tkp \to \infty$ as $k'\to \infty $ and $\tkpp \to \infty$ as $k'' \to \infty$, such that 
	\begin{align} \label{eq19}
	\dist(p(\tkp), \setp) > \delta, \quad \dist(p(\tkpp), \setc) > \delta.
	\end{align} 
	Since $p(t)$ converges to $\setm$, there exists $T>0$ such that $\forall t \ge T$, 
	\begin{equation} \label{eq20}
	\begin{split}
	&\dist(p(t), \setm) \le \gamma \implies 
	\\ &\min\{ \dist(p(t), \setc), \dist(p(t), \setp)\} \le \gamma,
	\end{split}
	\end{equation}
	where $0 < \gamma < \min\{\delta, \beta/4\}$. Due to \eqref{eq19} and \eqref{eq20}, it follows that 
	\begin{align} \label{eq19_2}
	\dist(p(\tkp), \setc) \le \gamma, \quad \dist(p(\tkpp), \setp) \le \gamma
	\end{align} 
	for all $\tkp \ge T$ and $\tkpp \ge T$. We can pick a point $p(t_m)$ from $\big( p(\tkp) \big)_{k'=1}^{\infty}$ and a point $p(t_n)$ from $\big( p(\tkpp) \big)_{k''=1}^{\infty}$ such that $t_n > t_m \ge T$ (see Fig. \ref{fig:lemma_convergepc}). Therefore, we have $\dist(p(t_m), \setc) \le \gamma<\delta$ (due to \eqref{eq19_2}) and $\dist(p(t_n), \setc)>\delta$ (due to \eqref{eq19}). Similarly, $\dist(p(t_m), \setp) > \delta$ (due to \eqref{eq19}) and $\dist(p(t_n), \setp) \le \gamma<\delta$ (due to \eqref{eq19_2}). By the continuity of $p(t)$ and Lemma \ref{lemma_triangle}, there exists $\tau \in (t_m, t_n)$ such that 
	\begin{equation} \label{eq21}
	\dist(p(\tau), \setp) = \dist(p(\tau), \setc) \ge \beta/2.
	\end{equation}
	Equation \eqref{eq21} implies that $\dist(p(\tau), \setm) \ge \beta/2 > \gamma$, contradicting \eqref{eq20}. Therefore, $p(t)$ indeed converges to $\setp$ or $\setc$. Combining the two arguments, we have proven that $p(t)$ converges to $\setp$ or $\setc$ exclusively. 	\hfill$\qed$
\end{pf}
\begin{rem}
	If $\dist(\setp, \setc)=0$, then $\beta=0$ in the preceding proof. In this case, it is possible that $p(t)$ converges neither to $\setp$ nor to $\setc$.
\end{rem}

\bibliographystyle{plain}
\bibliography{ref}   

\end{document}